\documentclass[11pt]{article}
\newcommand{\beq}{\begin{equation}}
\newcommand{\eeq}{\end{equation}}
\def\s{{\,\rm s}}
\def\g{{\,\rm g}}
\def\eV{\,{\rm eV}}
\def\keV{\,{\rm keV}}
\def\MeV{\,{\rm MeV}}
\def\GeV{\,{\rm GeV}}
\def\TeV{\,{\rm TeV}}
\def\sv{\left<\sigma v\right>}
\def\({\left(}
\def\){\right)}
\def\cm{{\,\rm cm}}
\def\K{{\,\rm K}}
\title{Dark matter from stable charged particles?}
\author{Maxim Yu. Khlopov\\Center for Cosmoparticle Physics "Cosmion"\\
Miusskaya Pl. 4, 125047, Moscow,
Russia\\
Moscow Engineering Physics Institute, 115409 Moscow, Russia, and \\
APC laboratory 10, rue Alice Domon et L\'eonie Duquet 75205 Paris
Cedex 13, France\\Maxim.Khlopov@roma1.infn.it}

\begin{document}
\maketitle

\begin{abstract}
Particle physics candidates for cosmological dark matter are usually considered as neutral and weakly interacting. However stable charged leptons and quarks can also exist and, hidden in elusive atoms, play the role of dark matter. The necessary condition for such scenario is absence of stable particles with charge -1 and effective mechanism for  suppression of free positively charged heavy species. These conditions are realized in several recently developed scenarios. In scenario based on Walking Technicolor model excess of stable particles with charge -2 and the corresponding dark matter density is naturally related with the value and sign of cosmological baryon asymmetry. The excessive charged particles are bound with primordial helium in techni-O-helium "atoms", maintaining specific nuclear-interacting form of dark matter. Some properties of techni-O-helium Universe are discussed.
\end{abstract}

\section{Introduction}
The modern theory of Universe, based on General Relativity, has evolved the triumph of Einstein's ideas by putting cosmological term, first introduced by A.Einstein in 1917 \cite{Lambda}, in the "standard" $\Lambda$CDM model. The corresponding dark energy is the dominant element of the modern Universe, maintaining $70\%$ of its total density.

 General Relativity and Dark Energy maintain the frame for the portrait of elementary particles in the Universe. To survive in the Universe the particles should be stable, as are nuclei and electrons, composing the visible matter.
However, one must also explain the modern dark matter
density, corresponding to $25\%$ of total density and exceeding the baryonic matter density by a factor of 5. The widely shared belief is that dark matter is nonbaryonic and consists of new stable particles.

For a particle with the
mass $m$ the particle physics time scale is $t \sim 1/m$ (here and further, if not indicated otherwise,
we use the units $\hbar = c = k = 1$), so in
particle world we refer to particles with lifetime $\tau \gg 1/m$
as to metastable. To be of cosmological significance metastable
particle should survive after the temperature of the Universe $T$
fell down below $T \sim m$, what means that the particle lifetime
should exceed $t \sim (m_{Pl}/m) \cdot (1/m)$. Such a long
lifetime should find reason in the existence of an (approximate)
symmetry. From this viewpoint, cosmology of dark matter is sensitive to the most
fundamental properties of microworld, to the conservation laws
reflecting strict or nearly strict symmetries of particle theory.

One can formulate the set of conditions under which new particles
can be considered as candidates to dark matter (see e.g. \cite{book,Cosmoarcheology,Bled06,Bled07} for review and reference).

 \begin{itemize}
\item[$\bullet$] The particles should be stable or have lifetime larger,  than age of the Universe. \item[$\bullet$] They should fit the measured density of dark matter. Effects of their decay or annihilation should be compatible with the observed fluxes of electromagnetic background radiation and cosmic rays. \item[$\bullet$] More complicated forms of scalar fields, primordial black holes and even evolved primordial large scale structures are also possible, but in the latter case the contribution to the total density is restricted by the condition of the observed homogeneity and isotropy of the Universe\item[$\bullet$] The candidates for dark matter should decouple from plasma and radiation at least before the beginning of matter dominated stage. This is necessary to provide formation of large scale structure at the observed level of anisotropy of the cosmic microwave background radiation. 
\end{itemize}
The easiest way to satisfy these conditions is to involve neutral weakly interacting particles.
However it is not the only particle physics solution for the dark matter problem. As we show here,
new stable particles can have electric charge, but effectively behave as neutral and sufficiently weakly interacting. 

Recently several elementary particle frames for heavy
stable charged particles were proposed:  \begin{itemize} \item[(a)]  A heavy quark and
heavy neutral lepton (neutrino with mass above half the Z-Boson
mass) of fourth generation \cite{N,Legonkov}; which can avoid
experimental constraints \cite{Q,Okun} and form composite dark
matter species \cite{I,lom,KPS06,Khlopov:2006dk}; \item[(b)] A Glashow's ``sinister''
heavy tera-quark $U$ and tera-electron $E$, which can form a tower
of tera-hadronic and tera-atomic bound states with ``tera-helium
atoms'' $(UUUEE)$ considered as dominant dark matter
\cite{Glashow,Fargion:2005xz}. \item[(c)] AC-leptons, predicted
in the extension \cite{5} of standard model, based on the approach
of almost-commutative geometry \cite{bookAC}, can form evanescent
AC-atoms, playing the role of dark matter
\cite{5,FKS,Khlopov:2006uv,Khlopov:2006dk}.\end{itemize}

Finally, it was recently shown in \cite{KK} that an elegant solution is possible in the
framework of walking technicolor models
\cite{Sannino:2004qp,Hong:2004td,Dietrich:2005jn,Dietrich:2005wk,Gudnason:2006ug,Gudnason:2006yj}
and can be realized without an {\it ad hoc} assumption on charged
particle excess, made in the approaches (a)-(c).

In all these models, predicting stable charged particles, the
particles escape experimental discovery, because they are hidden
in elusive atoms, maintaining dark matter of the modern Universe.
It offers new solution for the physical nature of the cosmological
dark matter.

This approach differs from the idea of dark matter, composed of
primordial bound systems of super heavy charged particles and
antiparticles, proposed earlier to explain the origin of Ultra
High Energy Cosmic Rays (UHECR) \cite{UHECR}. To survive to the
present time and to be simultaneously the source of UHECR
super heavy particles should satisfy a set of constraints, which in
particular exclude the possibility that they possess gauge charges
of the standard model.

The particles, considered here, participate in the standard model
interactions and we discuss the problems, related with various scenarios
of composite atom-like dark matter, formed by
heavy electrically charged stable particles.
\section{Charged components of composite dark matter}
\subsection{Charged tera-particles}
In Glashow's  "Sinister" $SU(3)_c \times SU(2) \times SU(2)'
\times U(1)$ gauge model \cite{Glashow} three heavy generations of
tera-fermions are related with the light fermions by $CP'$
transformation linking light fermions to charge conjugates of
their heavy partners and vice versa. $CP'$ symmetry breaking makes
tera-fermions much heavier than their light partners. Tera-fermion
mass pattern is the same as for light generations, but all the
masses are multiplied by the same factor $\sim 10^6$. It gives the
masses of lightest heavy particles in {\it tera}-eV (TeV) range,
explaining their name.  
Strict conservation of $F = (B-L) - (B' - L')$ prevents mixing of
charged tera-fermions with light quarks and leptons. Tera-fermions
are sterile relative to $SU(2)$ electroweak interaction, and do
not contribute into standard model parameters.
 In such realization the new heavy neutrinos ($N_i$)
  acquire large masses and their
mixing with light neutrinos $\nu$ provides a "see-saw" mechanism
of light neutrino Dirac mass generation. Here in a Sinister model
the heavy neutrino is unstable. On the contrary in this scheme
$E^-$ is the lightest heavy fermion and it is absolutely stable.
%%%%%%%%%%%%%%%%%%%%%%%%%%%%%%%%%%%%%%%%%%%%%%%%%%%%%%

Since the lightest quark $U$ of heavy generation does not mix with
quarks of 3 light generation, it can decay only to heavy
generation leptons owing to GUT-type interactions, what makes it
sufficiently long living. If its lifetime exceeds the age of the
Universe, primordial $U$-quark hadrons as well as heavy leptons
$E^-$ %%%%%%%%%%%%%%%%%%%%%%%%%%%%%%%%%%%%%%%%%%%%%%%%%%%%%%
should be present in the modern matter.

%Indeed,
%%%%%%%%%%%%%%%%%%%%%%%%%%%%%%%%%%%%%%%%%%%%%%%%%%%%%%
%in the
Glashow's "Sinister" scenario \cite{Glashow} took into account
that
%%%%%%%%%%%%%%%%%%%%%%%%%%%%%%%%%%%%%%%%%%%%%%%%%%%%%%
 very heavy quarks $Q$ (or antiquarks $\bar Q$) can form bound states with other heavy quarks
 (or antiquarks) due to their Coulomb-like QCD attraction, and the binding energy of these states
 substantially exceeds the binding energy of QCD confinement.
Then $(QQq)$ and $(QQQ)$ baryons can exist.

%In the model the properties of heavy generation fermions were
%fixed by their discrete $CP'$ symmetry with light fermions.
According to  \cite{Glashow} primordial heavy quark $U$ and heavy
electron $E$ are stable and
%%%%%%%%%%%%%%%%%%%%%%%%%%%%%%%%%%%%%%%%%%%%%%%%%%%%%%
may form a neutral most probable and stable (while being
evanescent) $(UUUEE)$ "atom"
%%%%%%%%%%%%%%%%%%%%%%%%%%%%%%%%%%%%%%%%%%%%%%%%%%%%%%
with $(UUU)$ hadron as nucleus and two $E^-$s as "electrons". The
tera gas of such "atoms" seemed an ideal candidate for a very new
and fascinating WIMP-like dark matter. \subsection{Stable AC leptons from almost commutative
geometry} The AC-model \cite{5} appeared as realistic elementary
particle model, based on the specific approach of \cite{bookAC} to
unify general relativity, quantum mechanics and gauge symmetry.

This realization naturally embeds the Standard model, both
reproducing its gauge symmetry and Higgs mechanism, but to be
realistic, it should go beyond the standard model and offer
candidates for dark matter. Postulates of noncommutative geometry
put severe constraints on the gauge symmetry group, excluding in
this approach, supersymmetric and GUT extensions as well as the extensive phenomenology
of superstrings. The
AC-model \cite{5} extends the fermion content of the Standard
model by two heavy particles with opposite electromagnetic and
Z-boson charges. Having no other gauge charges of Standard model,
these particles (AC-fermions) behave as heavy stable leptons with
charges $-2e$ and $+2e$, called here $A$ and $C$, respectively.
AC-fermions are sterile relative to $SU(2)$ electro-weak
interaction, and do not contribute to the standard model
parameters. The mass of AC-fermions is originated from
noncommutative geometry of the internal space (thus being much
less than the Planck scale) and is not related to the Higgs
mechanism. The lower limit ($\ge 100{\GeV} $) for this mass follows from absence of
new charged leptons in LEP. In the absence of AC-fermion
mixing with light fermions, AC-fermions can be absolutely stable.

The mechanism of baryosynthesis in the present
version of AC model is not clear, therefore the AC-lepton excess
was postulated in \cite{5,FKS,Khlopov:2006uv} to saturate the
modern CDM density (similar to the approach sinister model).
 Primordial excessive negatively charged $A^{--}$ and positively
charged $C^{++}$ form a neutral most probable and stable (while
being evanescent) $(AC)$ "atom", the AC-gas of such "atoms" being
a candidate for dark matter,
accompanied by small ($\sim 10^{-8}$) fraction of $^4HeA^{--}$ atoms
called in this case OLe-helium \cite{5,FKS,Khlopov:2006uv,Khlopov:2006dk}.

\subsection{Stable pieces of 4th generation matter} Precision data
on Standard model parameters admit \cite{Okun} the existence of
4th generation, if 4th neutrino ($N$) has mass about 50 GeV, while
masses of other 4th generation particles are close to their
experimental lower limits, being $>100$GeV for charged lepton
($E$) and $>300$GeV for 4th generation $U$ and $D$ quarks
\cite{CDF}. 

4th generation can follow from heterotic string phenomenology and
its difference from the three known light generations can be
explained by a new conserved charge, possessed only by
its quarks and leptons \cite{N,Q,I}. Strict conservation of this charge makes the
lightest particle of 4th family (4th neutrino $N$) absolutely
stable, while the lightest quark must be sufficiently long living
\cite{Q,I}. The lifetime of $U$ can exceed the age of the
Universe, as it was revealed in \cite{Q,I} for $m_U<m_D$.

$U$-quark can form lightest $(Uud)$ baryon and $(U \bar u)$ meson.
The corresponding antiparticles are formed by $\bar U$ with light
quarks and antiquarks. Owing to large chromo-Coulomb binding
energy ($\propto \alpha_{c}^2 \cdot m_U$, where $\alpha_{c}$ is
the QCD constant) stable double and triple $U$ bound states
$(UUq)$, $(UUU)$ and their antiparticles $(\bar U \bar U \bar u)$,
$(\bar U \bar U \bar U)$ can exist
\cite{Q,Glashow,Fargion:2005xz}. Formation of these double and
triple states at accelerators and in
cosmic rays is strongly suppressed, but they can form in early
Universe and strongly influence cosmological evolution of 4th
generation hadrons. As shown in \cite{I}, \underline{an}ti-
\underline{U}-\underline{t}riple state called \underline{anut}ium
or $\Delta^{--}_{3 \bar U}$ is of special interest. This stable
anti-$\Delta$-isobar, composed of $\bar U$ antiquarks is bound with $^4He$ in atom-like state
of O-helium \cite{I} or ANO-helium \cite{lom,KPS06,Khlopov:2006dk},
proposed as dominant forms of dark matter.
%%%%%%%%%%%%%%%%%%%%%%%%%%%%%%%%%%%%%%%%%%%%%%%%%%%%%%%%%%%%%%%%%%%%%%%%%%%%%%%%%%%%%%%%%%%%%%%%%
\subsection{Problems of composite dark matter scenarios}
In all these recent models, the predicted stable charged particles
escape experimental discovery, because they are hidden in elusive
atoms, composing the dark matter of the modern Universe. It offers
a new solution for the physical nature of the cosmological dark
matter. As it was recently shown in \cite{KK} such a solution is also possible in the
framework of walking technicolor models
\cite{Sannino:2004qp,Hong:2004td,Dietrich:2005jn,Dietrich:2005wk,Gudnason:2006ug,Gudnason:2006yj}
and can be realized without an {\it ad hoc} assumption on charged
particle excess, made in the approaches (a)-(c).

The approaches (b) and (c) try to escape the problems of free
charged dark matter particles \cite{Dimopoulos:1989hk} by hiding
opposite-charged particles in atom-like bound systems, which
interact weakly with baryonic matter. However, in the case of
charge symmetry, when primordial abundances of particles and
antiparticles are equal, annihilation in the early Universe
suppresses their concentration. If this primordial abundance still
permits these particles and antiparticles to be the dominant dark
matter, the explosive nature of such dark matter is ruled out by
constraints on the products of annihilation in the modern Universe
\cite{Q,FKS}. Even in the case of charge asymmetry with primordial
particle excess, when there is no annihilation in the modern
Universe, binding of positive and negative charge particles is
never complete and positively charged heavy species should retain.
Recombining with ordinary electrons, these heavy positive species
give rise to cosmological abundance of anomalous isotopes,
exceeding experimental upper limits. To satisfy these upper
limits, the anomalous isotope abundance on Earth should be
reduced, and the mechanisms for such a reduction are accompanied
by effects of energy release which are strongly constrained, in
particular, by the data from large volume detectors.

These problems of composite dark matter models \cite{Glashow,5}
revealed in \cite{Q,Fargion:2005xz,FKS,I}, can be avoided, if the
excess of only $-2$ charge $A^{--}$
%($(\bar U \bar U \bar u)$, $(\bar U \bar U \bar U)$)
% or neutral
%$(\bar U u)$
particles is generated in the early Universe. In
walking technicolor models, technilepton and technibaryon excess
is related to baryon excess and the excess of $-2$ charged
particles can appear naturally for a reasonable choice of model
parameters \cite{KK}. It distinguishes this case from other composite dark
matter models, since in all the previous realizations, starting
from \cite{Glashow}, such an excess was put by hand to saturate
the observed cold dark matter (CDM) density by composite dark
matter. Taking into account that the earlier scenarios were recently extensively
reviewed in \cite{Khlopov:2006dk}, we'll concentrate further on
the scenario \cite{KK}, based on the walking technicolor model.

\section{Dark Matter from Walking Technicolor}

The minimal walking technicolor model
\cite{Sannino:2004qp,Hong:2004td,Dietrich:2005jn,Dietrich:2005wk,Gudnason:2006ug,Gudnason:2006yj}
has two techniquarks, i.e. up $U$ and down $D$, that transform
under the adjoint representation of an $SU(2)$ technicolor gauge
group. The global symmetry of the model is an $SU(4)$ that breaks
spontaneously to an $SO(4)$. The chiral condensate of the
techniquarks breaks the electroweak symmetry. There are nine
Goldstone bosons emerging from the symmetry breaking. Three of
them are eaten by the $W$ and the $Z$ bosons. The remaining six
Goldstone bosons are $UU$, $UD$, $DD$ and their corresponding
antiparticles. These six
Goldstone bosons carry technibaryon number since they are made of
two techniquarks or two anti-techniquarks. This means that if no
processes violate the technibaryon number, the lightest
technibaryon will be stable. 

The electric charges of $UU$, $UD$,
and $DD$ are given in general by $y+1$, $y$, and $y-1$
respectively, where $y$ is an arbitrary real number. For any real
value of $y$, gauge anomalies are
canceled~\cite{Gudnason:2006yj}. The model requires in addition
the existence of a fourth family of leptons, i.e. a ``new
neutrino'' $\nu'$ and a ``new electron'' $\zeta$ in order to
cancel the Witten global anomaly. Their electric charges are in
terms of $y$ respectively $(1-3y)/2$ and $(-1-3y)/2$. The
effective theory of this minimal walking technicolor model has
been presented in~\cite{Gudnason:2006ug,Foadi:2007ue}.

There are several possibilities for a dark matter candidate
emerging from this minimal walking technicolor model. For the case
where $y=1$, the $D$ techniquark (and therefore also the $DD$
boson) become electrically neutral. If one assumes that $DD$ is
the lightest technibaryon, then it is absolutely stable, because
there is no way to violate the technibaryon number apart from the
sphalerons that freeze out close to the electroweak scale. This
scenario was studied in
Refs.~\cite{Gudnason:2006ug,Gudnason:2006yj}. 

 Within the same model and electric charge
assignment, there is another possibility. Since both techniquarks
and technigluons transform under the adjoint representation of the
$SU(2)$ group, it is possible to have bound states between a $D$
and a technigluon $G$. The object $D^{\alpha}G^{\alpha}$ (where
$\alpha$ denotes technicolor states) is techni-colorless. If such
an object has a Majorana mass, then it can account for the whole
dark matter density without being excluded by CDMS, due to the
fact that Majorana particles have no spin independent interaction with nuclei
and their non-coherent elastic cross section is very low for the
current sensitivity of detectors~\cite{Kouvaris:2007iq}. 

Finally, if one choose $y=1/3$, $\nu'$ has zero electric charge.
In this case the heavy fourth Majorana neutrino $\nu'$ can play
the role of a dark matter particle. This scenario was explored
first in~\cite{Kainulainen:2006wq} and later
in~\cite{Kouvaris:2007iq}. It was shown that indeed the fourth
heavy neutrino can provide the dark matter density without being
excluded by CDMS~\cite{Cline} or any other experiment.

Scenario of composite dark matter corresponds mostly the first case
 mentioned above, that is $y=1$ and the Goldstone bosons $UU$,
$UD$, and $DD$ have electric charges 2, 1, and 0 respectively. In
addition for $y=1$, the electric charges of $\nu'$ and $\zeta$ are
respectively $-1$ and $-2$. There are three possibilities for a scenario where
stable particles with $-2$ electric charge have substantial relic
densities and can capture $^4He^{++}$ nuclei to form a neutral
atom. 

The first
one is to have a relic density of $\bar{U}\bar{U}$, which has $-2$
charge. For this to be true we should assume that $UU$ is lighter
than $UD$ and $DD$ and no processes (apart from electroweak
sphalerons) violate the technibaryon number. The second one is to
have abundance of $\zeta$ that again has $-2$ charge and the third
case is to have both $\bar{U}\bar{U}$ (or $DD$ or
$\bar{D}\bar{D}$) and $\zeta$. 

For the first case to be realized,
$UU$ although charged, should be lighter than both $UD$ and $DD$.
This can happen if one assumes that there is an isospin splitting
between $U$ and $D$. This is not hard to imagine since for the
same reason in QCD the charged proton is lighter than the neutral
neutron. Upon making this assumption, $UD$ and $DD$ will decay
through weak interactions to the lightest $UU$. The technibaryon
number $TB$ is conserved and therefore $UU$ (or $\bar{U}\bar{U}$) is
stable. 

Similarly in the second case where $\zeta$ is the abundant
$-2$ charge particle, $\zeta$ must be lighter than $\nu'$ and
there should be no mixing between the fourth family of leptons and
the other three of the Standard Model. The technilepton number $L'$ number is violated
only by sphalerons and therefore after the temperature falls
roughly below the electroweak scale $\Lambda_{EW}$ and the
sphalerons freeze out, $L'$ is conserved, which means that the
lightest particle, that is $\zeta$ in this case, is absolutely
stable. It was also assumed in \cite{KK} that technibaryons decay to Standard Model
particles through Extended Technicolor (ETC) interactions and
therefore $TB=0$. 

Finally there is a
possibility to have both the $L'$ and $TB$ conserved after sphalerons
have frozen out. In this case, the dark matter would be composed
of bound atoms $(^4He^{++}\zeta^{--})$ and either $(^4He^{++}(\bar
U \bar U )^{--})$ or neutral $DD$ (or $\bar{D}\bar{D}$).
\section{\label{Interactions} The origin of Techni-O-helium}
 \subsection{Techniparticle excess}
The calculation of the  excess of the technibaryons with respect
to the one of the baryons
%for the Standard Model particles and
 was pioneered in
Refs.~\cite{Harvey:1990qw,Barr:1990ca,Chivukula,Khlebnikov:1996vj}. In \cite{KK} 
the excess of $\bar{U}\bar{U}$ and $\zeta$ was calculated 
along the lines of~\cite{Gudnason:2006yj}. The technicolor and the
Standard Model particles are in thermal equilibrium as long as the
rate of the weak (and color) interactions is larger than the
expansion of the Universe. In addition, the sphalerons allow the
violation $TB$, baryon number $B$, lepton number $L$, and $L'$ as
long as the temperature of the Universe is higher than roughly
$\Lambda_{EW}$. It is possible through the equations of thermal
equilibrium, sphalerons and overall electric neutrality for the
particles of the Universe, to associate the chemical potentials of
the various particles. The relationship between these chemical potentials 
with proper account for statistical factors, $\sigma$, results in relationship 
between $TB$, $B$, $L$, and $L'$ after sphaleron processes are frozen out \cite{KK}
\beq
\frac{TB}{B}=-\sigma_{UU}\left(\frac{L'}{B}\frac{1}{3\sigma_{\zeta}}+1+\frac{L}{3B}\right).
\label{tbb}\eeq 
Here $\sigma_i$ ($i=UU, \zeta$) are statistical factors.
It was shown in \cite{KK} that there can be excess of techni(anti)baryons, $(\bar{U}\bar{U})^{--}$,
technileptons $\zeta^{--}$ or of the both and parameters of model were found at which this asymmetry has
proper sign and value, saturating the dark matter density at the observed baryon asymmetry 
of the Universe.
\subsection{\label{Formation} Formation of techni-O-helium}
In the Standard Big Bang nucleosynthesis (SBBN), $^4He$ is formed with an
abundance $r_{He} = 0.1 r_B = 8 \cdot 10^{-12}$ and, being in
excess, binds all the negatively charged techni-species into
atom-like systems.

At a temperature $T<I_{o} = Z_{TC}^2 Z_{He}^2 \alpha^2 m_{He}/2
\approx 1.6 \MeV,$ where $\alpha$ is the fine structure constant,
and $Z_{TC}=-2$ stands for the electric charge of $\bar{U}\bar{U}$
and/or of $\zeta$, the reaction \beq
\zeta^{--}+^4He^{++}\rightarrow \gamma +(^4He\zeta) \label{EHeIg}
\eeq and/or \beq (\bar{U}\bar{U})^{--}+^4He^{++}\rightarrow \gamma
+(^4He(\bar{U}\bar{U})) \label{EHeIgU} \eeq can take place. Actually they can go only
after $^4He$ is formed in SBBN at $T \le 100 \keV$. In
these reactions neutral techni-O-helium ``atoms" are produced at $T_o \approx 60 \keV$ \cite{KK}. The
size of these ``atoms" is \cite{I,FKS,KK} \beq R_{o} \sim 1/(Z_{TC}
Z_{He}\alpha m_{He}) \approx 2 \cdot 10^{-13} \cm. \label{REHe}
\eeq
Virtually all the free $(\bar{U}\bar{U})$ and/or $\zeta$ (which will be further denoted by
$A^{--}$) are trapped by helium and their remaining abundance
becomes exponentially small.

For particles $Q^-$ with charge $-1$, as for tera-electrons in the
sinister model \cite{Glashow}, $^4He$ trapping results in the
formation of a positively charged ion $(^4He^{++} Q^-)^+$, result
in dramatic over-production of anomalous hydrogen
\cite{Fargion:2005xz}. Therefore, only the choice of $- 2$
electric charge for stable techniparticles makes it possible to
avoid this problem. In this case, $^4He$ trapping leads to the
formation of neutral {\it techni-O-helium} ``atoms"
$(^4He^{++}A^{--})$.

The formation of techni-O-helium reserves a fraction of $^4He$ and
thus it changes the primordial abundance of $^4He$. For the
lightest possible masses of the techniparticles $m_{\zeta} \sim
m_{TB} \sim 100 \GeV$, this effect can reach 50\% of the $^4He$
abundance formed in SBBN. Even if the mass of the techniparticles
is of the order of TeV, $5\%$ of the $^4He$ abundance is hidden in
the techni-O-helium atoms. This can lead to important consequences
once we compare the SBBN theoretical predictions to observations.

\subsection{Techni-O-helium in Big bang Nucleosynthesis}
The question of the participation of techni-O-helium in nuclear
transformations and its direct influence on the chemical element
production is less evident. Indeed, techni-O-helium looks like an
$\alpha$ particle with a shielded electric charge. It can closely
approach nuclei due to the absence of a Coulomb barrier. Because
of this, in the presence of techni-O-helium, the
character of SBBN processes should change drastically. However, it
might not lead to immediate contradiction with observations.

The following simple argument \cite{FKS,KK} can be used for suppression of 
binding of $A^{--}$ with nuclei heavier than $^4He$. In fact,
the size of techni-O-helium is of the order of the size of $^4He$
and for a nucleus $^A_ZQ$ with electric charge $Z>2$, the size of
the Bohr orbit for an $Q A^{--}$ ion is less than the size of the
nucleus $^A_ZQ$ (see \cite{Cahn}). This means that while binding with a heavy
nucleus, $A^{--}$ penetrates it and interacts effectively with a
part of the nucleus of a size less than the corresponding Bohr
orbit. This size corresponds to the size of $^4He$, making
techni-O-helium the most bound $Q A^{--}$ atomic state. It favors
a picture, according to which a techni-O-helium collision with a
nucleus, results in the formation of techni-O-helium and the whole
process looks like an elastic collision.

The interaction of the $^4He$ component of $(He^{++}A^{--})$ with
a $^A_ZQ$ nucleus can lead to a nuclear transformation due to the
reaction \beq ^A_ZQ+(HeA) \rightarrow ^{A+4}_{Z+2}Q +A^{--},
\label{EHeAZ} \eeq provided that the masses of the initial and
final nuclei satisfy the energy condition \beq M(A,Z) + M(4,2) -
I_{o}> M(A+4,Z+2), \label{MEHeAZ} \eeq where $I_{o} = 1.6 \MeV$ is
the binding energy of techni-O-helium and $M(4,2)$ is the mass of
the $^4He$ nucleus.

This condition is not valid for stable nuclei participating in
reactions of the SBBN. However, tritium $^3H$, which is also
formed in SBBN with abundance $^3H/H \sim 10^{-7}$ satisfies this
condition and can react with techni-O-helium, forming $^7Li$ and
opening the path of successive techni-O-helium catalyzed
transformations to heavy nuclei. This effect might strongly
influence the chemical evolution of matter on the pre-galactic
stage and needs a self-consistent consideration within the Big
Bang nucleosynthesis network. However, the following arguments \cite{FKS,KK}
show that this effect may not lead to immediate contradiction with
observations as it might be expected.
\begin{itemize}
\item[$\bullet$] On the path of reactions (\ref{EHeAZ}), the final
nucleus can be formed in the excited $(\alpha, M(A,Z))$ state,
which can rapidly experience an $\alpha$- decay, giving rise to
techni-O-helium regeneration and to an effective quasi-elastic
process of $(^4He^{++}A^{--})$-nucleus scattering. It leads to a
possible suppression of the techni-O-helium catalysis of nuclear
transformations. \item[$\bullet$] The path of reactions
(\ref{EHeAZ}) does not stop on $^7Li$ but goes further through
$^{11}B$, $^{15}N$, $^{19}F$, ... along the table of the chemical
elements. \item[$\bullet$] The cross section of reactions
(\ref{EHeAZ}) grows with the mass of the nucleus, making the
formation of the heavier elements more probable and moving the
main output away from a potentially dangerous Li and B
overproduction.
\end{itemize}
Such a qualitative change of the
physical picture appeals to necessity in a detailed nuclear
physics treatment of the ($A^{--}$+ nucleus) systems and of the
whole set of transformations induced by techni-O-helium. Though the above arguments
do not seem to make these dangers immediate and obvious, a
detailed study of this complicated problem is needed.

\section{\label{Oscenario} Techni-O-helium Universe}
\subsection{Gravitational instability of the techni-O-helium gas}
 Due to nuclear interactions of its helium constituent with
nuclei in cosmic plasma, the techni-O-helium gas is in thermal
equilibrium with plasma and radiation on the Radiation Dominance
(RD) stage, and the energy and momentum transfer from the plasma
is effective. The radiation pressure acting on plasma is then
effectively transferred to density fluctuations of techni-O-helium
gas and transforms them in acoustic waves at scales up to the size
of the horizon. However, as it was first noticed in \cite{I}, this
transfer to heavy nuclear-interacting species becomes ineffective
before the end of the RD stage and such species decouple from
plasma and radiation. Consequently, nothing prevents the
development of gravitational instability in the gas of these
species. This argument is completely applicable to the case of
techni-O-helium.

At temperature $T < T_{od} \approx 45 S^{2/3}_2\eV$, first
estimated in \cite{I} for the case of OLe-helium, the energy and
momentum transfer from baryons to techni-O-helium is not effective
because $n_B \sv (m_p/m_o) t < 1$, where $m_o$ is the mass of the
$tOHe$ atom and $S_2=\frac{m_o}{100 \GeV}$. Here \beq \sigma
\approx \sigma_{o} \sim \pi R_{o}^2 \approx
10^{-25}\cm^2\label{sigOHe}, \eeq and $v = \sqrt{2T/m_p}$ is the
baryon thermal velocity. The techni-O-helium gas decouples from
the plasma and plays the role of dark matter, which starts to
dominate in the Universe at $T_{RM}=1 \eV$.

The development of gravitational instabilities of the
techni-O-helium gas triggers large scale structure formation, and
the composite nature of techni-O-helium makes it more close to
warm dark matter.

The total mass of the $tOHe$ gas with density $\rho_d =
\frac{T_{RM}}{T_{od}} \rho_{tot}$ within the cosmological horizon
$l_h=t$ is
$$M=\frac{4 \pi}{3} \rho_d t^3.$$ In the period of decoupling $T = T_{od}$, this mass  depends
strongly on the techniparticle mass $S_2$ and is given by \beq
M_{od} = \frac{T_{RM}}{T_{od}} m_{Pl} (\frac{m_{Pl}}{T_{od}})^2
\approx 2 \cdot 10^{46} S^{-8/3}_2 \g = 10^{13} S^{-8/3}_2
M_{\odot}, \label{MEPm} \eeq where $M_{\odot}$ is the solar mass.
The techni-O-helium is formed only at $T_o \approx 60 \keV$ and its total mass
within the cosmological horizon in the period of its creation is
$$M_{o}=\frac{T_{RM}}{T_{o}} m_{Pl} (\frac{m_{Pl}}{T_{o}})^2=M_{od}\left(\frac{T_{o}}{T_{od}}\right)^3 = 10^{37} \g$$.

On the RD stage before decoupling, the Jeans length $\lambda_J$ of
the $tOHe$ gas was of the order of the cosmological horizon
$\lambda_J \sim l_h \sim t.$ After decoupling at $T = T_{od}$, it
falls down to $\lambda_J \sim v_o t,$ where $v_o =
\sqrt{2T_{od}/m_o}.$ Though after decoupling the Jeans mass in the
$tOHe$ gas correspondingly falls down  \cite{I,KK}
$$M_J \sim v_o^3 M_{od}\sim 3 \cdot 10^{-14}M_{od},$$ one should
expect strong suppression of fluctuations on scales $M<M_o$, as
well as adiabatic damping of sound waves in the RD plasma for
scales $M_o<M<M_{od}$. It provides suppression of small scale
structure in the considered model for all reasonable masses of
techniparticles.

The cross section of mutual collisions of techni-O-helium
``atoms'' is given by Eq.~(\ref{sigOHe}). The $tOHe$ ``atoms" can
be considered as collision-less gas in clouds with a number
density $n_{o}$ and a size $R$, if $n_{o}R < 1/\sigma_{o}$. This
condition is valid for the techni-O-helium gas in galaxies.

Mutual collisions of techni-O-helium ``atoms'' determine the
evolution timescale for a gravitationally bound system of
collision-less $tOHe$ gas $$t_{ev} = 1/(n \sigma_{o} v) \approx 2
\cdot 10^{20} (1 \cm^{-3}/n)^{7/6}\s,$$ where the relative
velocity $v = \sqrt{G M/R}$ is taken for a cloud of mass $M_o$ and
an internal number density $n$. This timescale exceeds
substantially the age of the Universe and the internal evolution
of techni-O-helium clouds cannot lead to the formation of dense
objects. Being decoupled from baryonic matter, the $tOHe$ gas does
not follow the formation of baryonic astrophysical objects (stars,
planets, molecular clouds...) and forms dark matter halos of
galaxies.

\subsection{Techniparticle component of cosmic rays}
 The nuclear interaction of techni-O-helium with cosmic rays gives rise to
ionization of this bound state in the interstellar gas and to
acceleration of free techniparticles in the Galaxy. During the
lifetime of the Galaxy $t_G \approx 3 \cdot 10^{17} \s$, the
integral flux of cosmic rays $$F(E>E_0)\approx 1 \cdot
\left(\frac{E_0}{1 \GeV} \right)^{-1.7} \cm^{-2} \s^{-1}$$ can
disrupt the fraction of galactic techni-O-helium $ \sim
F(E>E_{min}) \sigma_o t_G \le 10^{-3},$ where we took $E_{min}
\sim I_o.$ Assuming a universal mechanism of cosmic ray
acceleration, a universal form of their spectrum, taking into
account that the $^4He$ component corresponds to $\sim 5$\% of the
proton spectrum, and that the spectrum is usually reduced to the
energy per nucleon, the $-2$ charged techniparticle component with anomalously low $Z/A$
can be present in cosmic rays at a level
of \beq
 \frac{A^{--}}{He} \ge 3 \cdot 10^{-7} \cdot S_2^{-3.7}. \eeq
 This flux may be within the reach for PAMELA and AMS02 cosmic ray
experiments.

Recombination of free techniparticles with protons and nuclei in
the interstellar space can give rise to radiation in the range
from few tens of keV - 1 MeV. However such a radiation is below
the cosmic nonthermal electromagnetic background radiation
observed in this range.

\subsection{\label{EpMeffects} Effects of techni-O-helium catalyzed processes in the Earth}
The first evident consequence of the proposed excess is the
inevitable presence of $tOHe$ in terrestrial matter. This is
because terrestrial matter appears opaque to $tOHe$ and stores all
its in-falling flux.

If the $tOHe$ capture by nuclei is not effective, its diffusion in
matter is determined by elastic collisions, which have a transport
cross section per nucleon \beq \sigma_{tr} = \pi R_{o}^2
\frac{m_{p}}{m_o} \approx 10^{-27}/S_2 \cm^{2}. \label{sigpEpcap}
\eeq In atmosphere, with effective height $L_{atm} =10^6 \cm$ and
baryon number density $n_B= 6 \cdot 10^{20} \cm^{-3}$, the opacity
condition $n_B \sigma_{tr} L_{atm} = 6 \cdot 10^{-1}/S_2$ is not
strong enough. Therefore, the in-falling $tOHe$ particles are
effectively slowed down only after they fall down terrestrial
surface in $16 S_2$ meters of water (or $4 S_2$ meters of rock).
Then they drift with velocity $V = \frac{g}{n \sigma v} \approx 8
S_2 A^{1/2} \cm/\s$ (where $A \sim 30$ is the average atomic
weight in terrestrial surface matter, and $g=980~
\cm/\s^2$), sinking down the center of the Earth on a
timescale $t = R_E/V \approx 1.5 \cdot 10^7 S_2^{-1}\s$, where
$R_E$ is the radius of the Earth.

The in-falling techni-O-helium flux from dark matter halo is
$\mathcal{F}=n_o v_h/8 \pi$, where the number density of $tOHe$ in
the vicinity of the Solar System is $n_o=3 \cdot 10^{-3}S_2^{-1}
\cm^{-3}$ and the averaged velocity $v_h \approx 3 \cdot
10^{7}\cm/\s$. During the lifetime of the Earth ($t_E \approx
10^{17} \s$), about $2 \cdot 10^{38}S_2^{-1}$ techni-O-helium
atoms were captured. If $tOHe$ dominantly sinks down the Earth, it
should be concentrated near the Earth's center within a radius
$R_{oc} \sim \sqrt{3T_c/(m_o 4 \pi G \rho_c)}$, which is $\le 10^8
S_2^{-1/2}\cm$, for the Earth's central temperature $T_c \sim 10^4
\K$ and density $\rho_c \sim 4 \g/\cm^{3}$.

Near the Earth's surface, the techni-O-helium abundance is
determined by the equilibrium between the in-falling and
down-drifting fluxes. It gives $$n_{oE}=2 \pi \mathcal{F}/V = 3
\cdot 10^3 \cdot S_2^{-2}\cdot A^{-1/2}\cm^{-3},$$ or for $A \sim
30$ about $5 \cdot 10^2 \cdot S_2^{-2} \cm^{-3}$. This number
density corresponds to the fraction $$f_{oE} \sim 5 \cdot 10^{-21}
\cdot S_2^{-2}$$ relative to the number density of the terrestrial
atoms $n_A \approx 10^{23} \cm^{-3}$.

These neutral $(^4He^{++}A^{--})$ ``atoms" may provide a catalysis
of cold nuclear reactions in ordinary matter (much more
effectively than muon catalysis). This effect needs a special and
thorough investigation. On the other hand, if $A^{--}$ capture by
nuclei, heavier than helium, is not effective and does not lead to
a copious production of anomalous isotopes, the
$(^4He^{++}A^{--})$ diffusion in matter is determined by the
elastic collision cross section (\ref{sigpEpcap}) and may
effectively hide techni-O-helium from observations.

One can give the following argument for an effective regeneration
and quasi-elastic collisions of techni-O-helium in terrestrial
matter. The techni-O-helium can be destroyed in the reactions
(\ref{EHeAZ}). Then, free $A^{--}$ are released and due to a
hybrid Auger effect (capture of $A^{--}$, ejection of ordinary $e$
from the atom with atomic number $A$, and charge of the nucleus
$Z$), $A^{--}$-atoms are formed, in which $A^{--}$ occupies highly
an excited level of the $(_Z^AQA)$ system, which is still much
deeper than the lowest electronic shell of the considered atom.
The $(_Z^AQA)$ atomic transitions to lower-lying states cause
radiation in the intermediate range between atomic and nuclear
transitions. In course of this falling down to the center of the
$(Z-A^{--})$ system, the nucleus approaches $A^{--}$. For $A>3$
the energy of the lowest state $n$ (given by $E_n=\frac{M \bar
\alpha^2}{2 n^2} = \frac{2 A m_p Z^2 \alpha^2}{n^2}$)  of the
$(Z-A^{--})$ system (having reduced mass $M \approx A m_p$) with a
Bohr orbit $r_n=\frac{n}{M \bar \alpha}= \frac{n}{2 A Z m_p
\alpha}$, exceeding the size of the nucleus $r_A \sim
A^{1/3}m^{-1}_{\pi}$ ($m_{\pi}$ being the mass of the pion), is
less than the binding energy of $tOHe$. Therefore the regeneration
of techni-O-helium in a reaction, inverse to (\ref{EHeAZ}), takes
place. An additional reason for the domination of the elastic
channel of the reactions (\ref{EHeAZ}) is that the final state
nucleus is created in the excited state and its de-excitation via
$\alpha$-decay can also result in techni-O-helium regeneration. If
regeneration is not effective and $A^{--}$ remains bound to the
heavy nucleus, anomalous isotope of $Z-2$ element should appear.
This is a serious problem for the considered model.

However, if the general picture of sinking down is valid, it might
give no more than the ratio $f_{oE} \sim 5 \cdot 10^{-21} \cdot S_2^{-2}$
of number density of anomalous isotopes to the number density of atoms of
terrestrial matter around us, which is below the experimental
upper limits for elements with $Z \ge 2$. For comparison, the best
upper limits on the anomalous helium were obtained in \cite{exp3}.
It was found, by searching with the use of laser spectroscopy for
a heavy helium isotope in the Earth's atmosphere, that in the mass
range 5 GeV - 10000 GeV, the terrestrial abundance (the ratio of
anomalous helium number to the total number of atoms in the Earth)
of anomalous helium is less than $2 \cdot 10^{-19}$ - $3 \cdot
10^{-19}$.

\subsection{\label{DMdirect} Direct search for techni-O-helium}

%%%%%%%%%%%%%%%%Insertion1%%%%%%%%%%%%%%%%%%%%%%%%%%%%%%%%%%%%%%%%%%%%%%%%%%%%%%%%%%%%
It should be noted that the nuclear cross section of the
techni-O-helium interaction with matter escapes the severe
constraints \cite{McGuire:2001qj} on strongly interacting dark
matter particles (SIMPs) \cite{Starkman,McGuire:2001qj} imposed by
the XQC experiment \cite{XQC}.
%%%%%%%%%%%%%%%%%%%%%%%%%%%%%%%%%%%%%%%%%%%%%%%%%%%%%%%%%%%%%%%%%%%%%%%%

In underground detectors, $tOHe$ ``atoms'' are slowed down to
thermal energies and give rise to energy transfer $\sim 2.5 \cdot
10^{-3} \eV A/S_2$, far below the threshold for direct dark matter
detection. It makes this form of dark matter insensitive to the
CDMS constraints. However, $tOHe$ induced nuclear transformation
can result in observable effects.

 Therefore, a special strategy of such a search is needed, that
can exploit sensitive dark matter detectors
%before they are
%installed underground.
 on the ground or in space. In particular, as it was revealed in
\cite{Belotsky:2006fa}, a few $\g$ of superfluid $^3He$ detector
\cite{Winkelmann:2005un}, situated in ground-based laboratory can
be used to put constraints on the in-falling techni-O-helium flux
from the galactic halo.
\section{\label{Discussion} Discussion}
To conclude, the existence of heavy stable particles is one of the popular solutions for dark matter problem. 
If stable particles have electric charge, dark matter candidates
can be atom-like states, in which negatively and positively 
charged particles are bound by Coulomb attraction. In this case
there is a serious problem to prevent overproduction of accompanying anomalous forms of
atomic matter.

Indeed, recombination of charged species is never complete in the
expanding Universe, and significant fraction of free charged
particles should remain unbound. Free positively charged species
behave as nuclei of anomalous isotopes, giving rise to a danger of
their over-production. Moreover, as soon as $^4He$ is formed in
Big Bang nucleosynthesis it captures all the free negatively
charged heavy particles. If the charge of such particles is -1 (as
it is the case for tera-electron in \cite{Glashow}) positively
charged ion $(^4He^{++}E^{-})^+$ puts Coulomb barrier for any
successive decrease of abundance of species, over-polluting modern
Universe by anomalous isotopes. It excludes the possibility of
composite dark matter with $-1$ charged constituents and only $-2$
charged constituents avoid these troubles, being trapped by helium
in neutral OLe-helium, O-helium (ANO-helium) or techni-O-helium states.

The existence of $-2$ charged states and the absence of stable
$-1$ charged constituents can take place in AC-model, in charge
asymmetric model of 4th generation and walking technicolor model with stable doubly charged technibaryons and/or technileptons.

To avoid overproduction of anomalous isotopes, an excess of $-2$
charged particles over their antiparticles should be
generated in the Universe. In all the earlier realizations of
composite dark matter scenario, this excess was put by hand to
saturate the observed dark matter density. In walking technicolor model this
abundance of -2 charged techibaryons and/or technileptons is connected
naturally to the value and sign of cosmological baryon asymmetry. These doubly charged $A^{--}$
techniparticles bind with $^4He$ in the
 neutral techni-O-helium states. For reasonable
values of the techniparticle mass, the amount of primordial $^4He$,
bound in this atom like state is significant and should be taken
into account in comparison to observations.

A challenging problem is the nuclear
transformations, catalyzed by techni-O-helium. The question about
their consistency with observations remains open, since special
nuclear physics analysis is needed to reveal what are the actual
techni-O-helium effects in SBBN and in terrestrial matter. However, qualitatively
one can expect a path for O-helium catalysis of heavy elements, making primordial 
heavy elements a signature for composite dark matter scenarios.

The destruction of techni-O-helium by cosmic rays in the Galaxy
releases free charged techniparticles, which can be accelerated
and contribute to the flux of cosmic rays. In this context, the
search for techniparticles at accelerators and in cosmic rays
acquires the meaning of a crucial test for the existence of the
basic components of the composite dark matter. At accelerators,
techniparticles would look like stable doubly charged heavy
leptons, while in cosmic rays, they represent a heavy $-2$ charge
component with anomalously low ratio of electric charge to mass.

The presented arguments enrich the class of possible particles, which can follow from
extensions of the Standard Model and be considered as dark matter candidates.
One can extend the generally accepted viewpoint that new stable particles should be neutral and weakly interacting.
We have seen that they can also be charged and play the role of dark matter because they are hidden in 
atom-like states, which are not the source of visible light. The constraints on such particles are very strict
and open a very narrow window for this new cosmologically interesting degree of freedom in particle theory. However, taking into account the exciting ability of O-helium to catalyze nuclear transformations of chemical elements, it is hard to estimate the expectation value of its discovery.  

\section*{Acknowledgements}
 I am grateful to co-authors K.Belotsky, C. Kouvaris, K.Shibaev and C. Stephan for fruitful collaboration in obtaining the presented results.

%\end{verbatim}

\end{document}